\documentclass[cameraready]{Interspeech}

\title{LoRA-Tuned Large Language Models for Dementia Detection via Multi-View Speech-Derived Features}

\author[affiliation={1}, orcid=0009-0003-3544-8040, equalcontribution]{Jonghyeon}{Park}
\author[affiliation={2}, orcid=0009-0003-6532-5068, equalcontribution]{Olivier Jiyoun}{Jung}
\author[affiliation={1}, orcid=0009-0007-7683-1117,correspondingauthor]{Myungwoo}{Oh}

\address{
    $^1$ NAVER Cloud, South Korea \\
    $^2$ Division of Communication and Media, Ewha Womans University, South Korea
}

\email{jong-hyeon.park@navercorp.com, olivierjiyounjung@gmail.com, myungwoo.oh@navercorp.com}


\usepackage{comment}

\begin{document}

\maketitle

\begin{abstract}
    Early detection of dementia enables timely intervention, and reflecting cognitive impairment, spontaneous speech offers a non-invasive screening modality. Conventional approaches often focus on a single representational dimension—such as acoustic descriptors, pause modeling, automatic speech recognition (ASR) transcripts, or multimodal fusion—limiting integrative reasoning across heterogeneous cognitive symptoms. We propose a low-rank adaptation (LoRA)-tuned large language model (LLM) that performs structured multi-view reasoning over four complementary speech-derived signals: ASR transcripts with pause markers, discourse-level topic cues, temporal fluency statistics, and phonological sequences. These cues are encoded within a unified prompt, enabling a single LLM to learn a coherent decision function without modality-specific encoders or late-stage fusion. On ADReSSo, our best model achieves an F1-score of 90.14\%, and ablation confirms the complementary contribution of each view.
\end{abstract}

\section{Introduction}

Dementia progressively impairs cognitive and communicative abilities, making early detection critical for timely intervention. Because these impairments often manifest in speech, spontaneous verbal output provides a rich source of behavioral markers reflecting cognitive status. Advances in machine learning have enabled automatic analysis of speech signals, making speech-based approaches promising and practical for early dementia detection. Recent surveys of speech-based Alzheimer's detection systems emphasize that dementia-related speech alterations span acoustic, temporal, and higher-level linguistic domains, with temporal and prosodic deviations repeatedly identified as among the most discriminative markers~\cite{Ding2024SpeechSurveyAD}.

One prominent way to capture the acoustic–prosodic dimension is through standardized feature sets such as eGeMAPS~\cite{Eyben2016GeMAPS}, extracted via openSMILE~\cite{Eyben2010openSMILE}. By aggregating carefully selected spectral, pitch, energy, and voice-quality descriptors into a compact representation, eGeMAPS provides a robust and interpretable baseline that has been widely adopted in speech-based dementia detection.

More recent work has leveraged data-driven speech representations. Self-supervised models such as Wav2vec-2.0~\cite{Baevski2020wav2vec2} and HuBERT~\cite{Hsu2021HuBERT} learn powerful acoustic embeddings directly from raw audio, substantially improving speech representation without manual feature engineering. However, while these models capture rich contextual acoustic patterns, they do not explicitly model temporal hesitation cues such as pause and speech duration.

Building on these representations, WavBERT~\cite{zhu21e_interspeech} integrates Wav2vec-based acoustic embeddings with bidirectional encoder representations from transformers (BERT)-based~\cite{devlin2019bert} textual representations and further incorporates connectionist temporal classification (CTC)~\cite{graves2006connectionist}-derived pause-duration features. This combination demonstrates that augmenting strong self-supervised speech embeddings with explicit temporal modeling yields improved performance for Alzheimer’s dementia detection. Phoneme-level representations have also been explored to model segmental variations in dementia speech, showing that incorporating phonological information can further enhance detection performance~\cite{SunQPLCZ25}.

Large-scale automatic speech recognition (ASR) models have also been used as transferable backbones. Whisper~\cite{radford2023robust}, a state-of-the-art ASR model, produces robust transcriptions for spontaneous speech and has become a strong backbone for clinical speech tasks. Improved transcription fidelity has been associated with stronger downstream dementia detection performance~\cite{Li24_whsiperbased}. Several studies have explored leveraging Whisper’s multilingual capability to extend dementia detection models beyond English, enabling cross-lingual clinical speech analysis and improving generalizability across languages~\cite{jia25_interspeech}.

With the emergence of large language models (LLMs), recent studies have explored their potential for transcript-level dementia detection. These approaches often leverage prompting strategies, such as prompt engineering and chain-of-thought prompting, to exploit the reasoning capabilities of LLMs~\cite{zheng2024alzheimer, park2025reasoning, ntampakis2025neuroxvocal}.

Despite these advances, most existing systems model these signals independently or combine them through separate encoders prior to aggregation. Such designs limit integrative reasoning, as dementia-related impairment manifests jointly across complementary speech dimensions.

In this work, we propose a framework that adapts an LLM via low-rank adaptation (LoRA)~\cite{hu2022lora} to jointly reason over complementary speech-derived representations encoded in a structured prompt. Lexical transcripts augmented with alignment-derived pause markers, temporal fluency indicators from forced alignment, phonological sequences from a disfluency-aware phoneme recognizer~\cite{guo2026huper}, and discourse-level topic and cluster cues are unified within a single input representation. Experiments on ADReSSo~\cite{luz21_interspeech} show that structured multi-view reasoning with parameter-efficient LLM adaptation provides a strong and unified framework for dementia detection.

Our contributions are summarized as follows:
\begin{itemize}
  \item We propose a unified multi-view speech-derived feature framework for dementia detection, leveraging a LoRA-tuned LLM to jointly reason over heterogeneous representations within a structured prompt.
  \item We show that integrating lexical, temporal, phonological, and discourse-level cues within a single reasoning framework yields consistent improvements over single-view and independently modeled baselines.
\end{itemize}

\section{Methods}
\label{sec:Methods}

\subsection{Overview}

We present a framework that trains an LLM using LoRA~\cite{hu2022lora} for speech-based dementia detection that integrates multi-view speech-derived features within a structured prompt. Our hypothesis is that dementia-related impairment manifests across complementary dimensions of speech, and that a single LLM can effectively learn to detect these cues when they are encoded within a unified representation.

\begin{figure}[t]
  \centering
  \includegraphics[width=\linewidth]{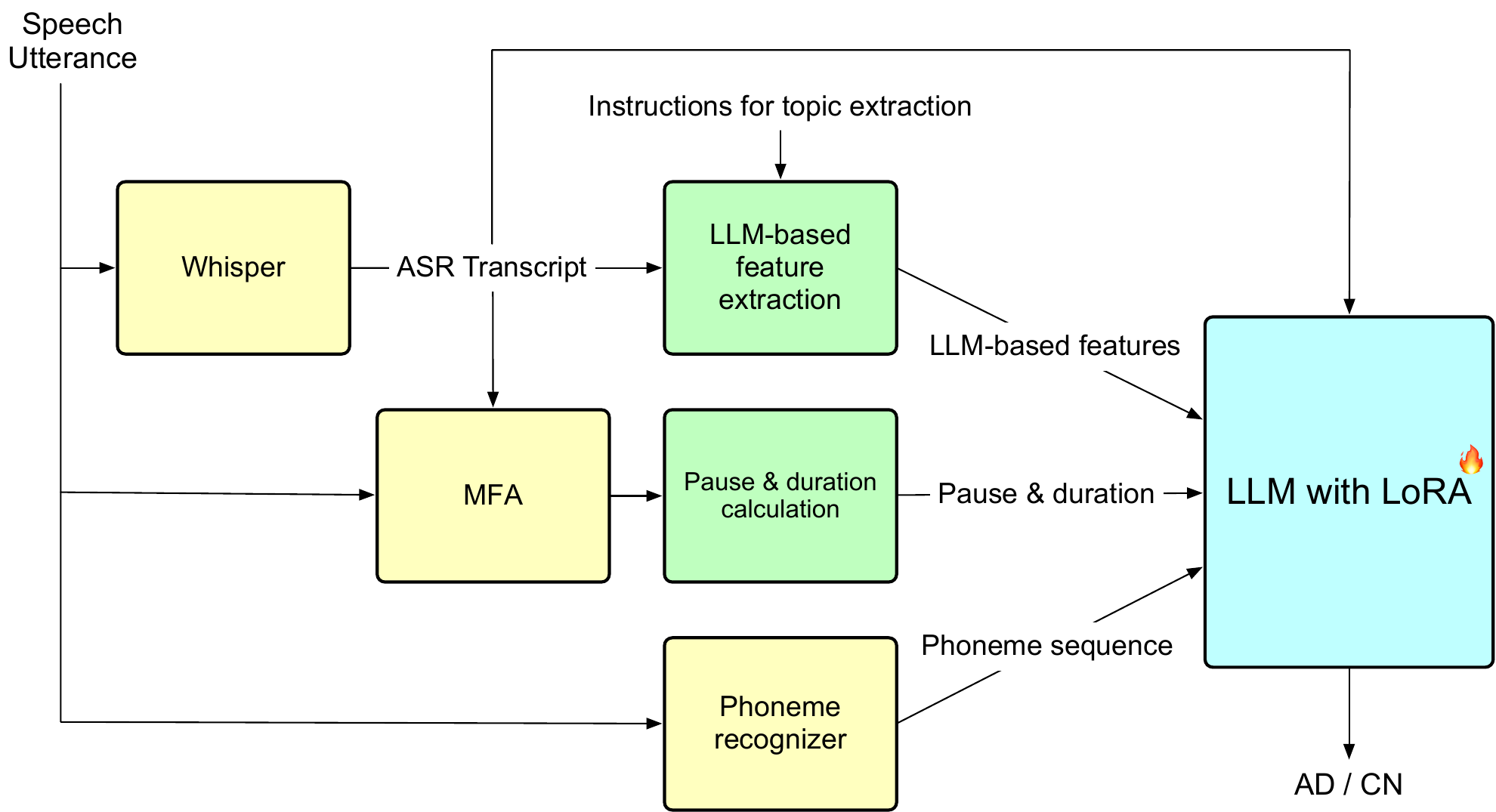}
  \caption{Schematic diagram of proposed pipeline.}
  \label{fig:overall_diagram}
\end{figure}

Instead of relying on separate models or conventional fusion pipelines, we incorporate heterogeneous speech-derived indicators into a structured prompt format. The LoRA-adapted LLM then performs reasoning over both sequential patterns and global summary features to produce a unified clinical prediction. An overview of the proposed pipeline is illustrated in Figure~\ref{fig:overall_diagram}.

\subsection{Multi-View Feature Extraction}
For each utterance, we derive four complementary speech-derived representations and encode them into a single structured input for LLM fine-tuning.

\subsubsection{Lexical Representation}
We generate lexical transcripts using Whisper~\cite{radford2023robust} large-v3, a transformer-based encoder-decoder ASR model pre-trained on large-scale weakly supervised audio data. Its robustness to spontaneous and disfluent speech is well suited for dementia corpora, where hesitations, repetitions, and incomplete utterances are common.

\subsubsection{Temporal Fluency Representation}

We derive temporal fluency indicators from word-level forced alignment using Montreal Forced Aligner (MFA)~\cite{mcAuliffe2017MFA}, which provides more reliable word boundaries than Whisper's word-level timestamps. From the aligned inter-word silence intervals, we compute utterance-level features, including speech rate (words per second), mean pause duration, and pause counts.

To determine the optimal pause threshold, we conduct an effect-size analysis comparing the Alzheimer’s disease (AD) and cognitively normal (CN) groups in the ADReSSo dataset~\cite{luz21_interspeech} (Table~\ref{tab:temporal_effect_size}). The 0.5-second threshold yields the strongest group separation (Cohen's $d=0.282$), while shorter thresholds reduce discriminative power as brief pauses are common in normal speech. Speech rate shows a comparatively modest effect ($d=-0.125$); however, it captures a complementary aspect of temporal fluency reflecting global speech tempo, whereas pause-related features primarily characterize hesitation patterns. We therefore adopt 0.5s as the operational threshold, applied consistently to both the numerical fluency features and the \texttt{<pause>} token insertion in the lexical stream.

\begin{table}[t]
  \centering
  \caption{Effect-size-based temporal fluency comparison between AD and CN groups.
  Pause duration is analyzed across clinically relevant thresholds.}
  \label{tab:temporal_effect_size}
  \resizebox{\columnwidth}{!}{%
  \begin{tabular}{lcccc}
  \toprule
  Metric & AD Mean & CN Mean & Diff & Cohen's $d$ \\
  \midrule
  \multicolumn{5}{c}{\textbf{Pause Threshold Analysis}} \\
  \midrule
  $\geq$ 0.3s & 1.2518 & 0.9459 & +0.3058 & 0.253 \\
  $\geq$ 0.5s & 1.5544 & 1.1828 & +0.3716 & 0.282 \\
  $\geq$ 1.0s & 2.2164 & 1.8476 & +0.3688 & 0.230 \\
  $\geq$ 1.5s & 2.9638 & 2.5946 & +0.3692 & 0.195 \\
  \midrule
  \multicolumn{5}{c}{\textbf{Speech Rate}} \\
  \midrule
  Words per second & 2.90 & 3.08 & $-$0.18 & $-$0.125 \\
  \bottomrule
  \end{tabular}%
  }
  \end{table}

\subsubsection{Discourse-Oriented Representation}

The ADReSSo dataset~\cite{luz21_interspeech} uses the Cookie Theft picture description task~\cite{goodglass1983boston}, in which all participants describe the same visual scene. This shared stimulus enables direct comparison of discourse patterns across speakers, making deviations in content coverage or narrative coherence interpretable as markers of cognitive impairment.

To capture discourse organization beyond surface lexical features, we define a fixed set of eight semantic clusters using a vision-capable LLM (GPT-5.2~\cite{openai2025gpt5}). The model identifies groupings that are (i) \emph{spatially grounded}, reflecting regions and objects in the scene; (ii) \emph{thematically coherent}, capturing related actions and entities; and (iii) \emph{discourse-aware}, including categories for meta-cognitive processes. Each cluster is further subdivided into fine-grained topics representing specific elements that speakers commonly mention.\footnote{The details are provided in https://github.com/vivivic/is26dementia.}

\begin{table}[t]
\centering
\caption{Discourse cluster scheme for Cookie Theft picture description.}
\label{tab:cluster_scheme}
\small
\resizebox{\columnwidth}{!}{
\begin{tabular}{@{}cl>{\raggedright\arraybackslash}p{5.5cm}@{}}
\toprule
\textbf{Cluster} & \textbf{Attentional Zone} & \textbf{Constituent Topics} \\
\midrule
C1 & Boy \& Cabinet Action & boy\_on\_stool, reaching\_up, cookie\_jar, taking\_cookie, stool\_balance, cabinet\_shelf, open\_cabinet \\
C2 & Girl Participation & girl\_reaching, asking\_for\_cookie, standing\_left, watching\_boy \\
C3 & Mother \& Domestic Activity & mother\_drying, holding\_plate, dish\_towel, ignoring\_children, kitchen\_apron \\
C4 & Water Overflow Event & overflowing\_sink, running\_faucet, water\_spilling, puddle\_on\_floor, water\_splash \\
C5 & Counter Objects & countertop\_items, plate\_on\_counter, cup\_and\_saucer, stacked\_dishes \\
C6 & Window \& Exterior & window\_view, curtains, outside\_scene \\
C7 & Room Layout \& Setting & kitchen\_setting, cabinets\_drawers, sink\_area, floor\_space \\
C8 & Meta-discourse & uncertainty, self\_correction, meta\_filler, retrieval\_difficulty, off\_topic, task\_management \\
\bottomrule
\end{tabular}
}
\end{table}

Utterance labeling is performed separately from defining the set of clusters. Given the fixed scheme, we assign each utterance a cluster and topic label via zero-shot inference with a text-based LLM~\cite{openai2025gpt5}, providing the transcript and cluster definitions as input. These labels are then encoded into the multi-view representation.

\subsubsection{Phonological Representation}

Phonological sequences are extracted using HuPER~\cite{guo2026huper}, a human-inspired phoneme recognizer designed to remain robust under degraded and disfluent speech. Unlike conventional phoneme extraction pipelines that rely primarily on bottom-up acoustic evidence, HuPER employs an adaptive inference strategy that leverages top-down linguistic constraints when acoustic cues are weak or ambiguous—conditions frequently observed in dementia speech.

In addition, HuPER’s WFST-based decoding further improves the stability of phoneme recognition under disfluency and articulation variability, producing consistent ARPAbet-style outputs. As a result, the extracted phonological sequences preserve articulatory and fluency-related information beyond lexical content, providing a complementary diagnostic signal for LLM-based dementia classification.

\subsection{Classification Model and Training}

We adopt several open-source LLMs such as Qwen3~\cite{yang2025qwen3} and Gemma-3~\cite{gemmateam2025gemma3technicalreport} and fine-tune via LoRA~\cite{hu2022lora} for utterance-level binary dementia classification. LoRA enables parameter-efficient adaptation of the large-scale model while preserving the pretrained representations, which is particularly important given the limited size of dementia speech corpora. The instruction-following capability of LLMs enables effective reasoning over the structured multi-view prompt, allowing the model to jointly interpret lexical, temporal, phonological, and discourse-level cues within a single inference pass.

\subsection{Structured Multi-View Prompt}

The four feature views are unified into a single JSON-structured prompt per utterance. As illustrated in Figure~\ref{fig:prompt_example}, each field directly corresponds to one representational level: \texttt{speech} carries the Whisper~\cite{radford2023robust} ASR transcript augmented with inline \texttt{<pause>} tokens; \texttt{cluster} and \texttt{topic} encode the discourse label derived from the LLM-based annotation pipeline; \texttt{num\_pause}, \texttt{mean\_pause\_sec}, and \texttt{words\_per\_second} summarize global temporal fluency; and \texttt{phoneme} provides the ARPAbet phoneme sequence from HuPER~\cite{guo2026huper}.

\begin{figure}[t]
\centering
{\scriptsize
\noindent\hrule
\vspace{-0.6em}
\begin{verbatim}
{
    "utterance": {
        "speech": "theres a pathway <pause> with ...
        "cluster": "C6",
        "topic": "window_curtains_outside",
        "num_pause": 1,
        "mean_pause_sec": 0.69,
        "words_per_second": 1.713,
        "phoneme": "EH ER Z DH AH P AE TH W EY W...
    }
}
\end{verbatim}
\vspace{-0.6em}
\hrule
}
\vspace{0.6em}
\caption{Example of structured multi-view prompt for a single utterance from the ADReSSo dataset. Each field encodes a distinct representational view.}
\label{fig:prompt_example}
\end{figure}

\section{Experiments}

\subsection{Datasets}

We evaluate our approach on the ADReSSo challenge dataset~\cite{luz21_interspeech}, which is a widely used benchmark for speech-based dementia detection. The dataset is derived from the DementiaBank Pitt corpus~\cite{becker1994natural} and is based on the Cookie Theft picture description task. ADReSSo is a transcript-free challenge supplying only raw audio with speaker-level binary labels for 237 participants (166 train, 71 test). The dataset features balanced class distributions to enable unbiased evaluation. For utterance segmentation, we use the speaker-turn boundaries provided by the ADReSSo challenge and restrict training and evaluation to participant turns only, discarding interviewer utterances. 

\subsection{Feature Implementation}

We extracted input features following the protocols in section ~\ref{sec:Methods}. Lexical transcripts were first generated using Whisper~\cite{radford2023robust} large-v3. These transcripts were then used as input to MFA~\cite{mcAuliffe2017MFA} to obtain word-level forced alignments. Based on the alignment results, silence intervals of $\ge$0.5s were identified and encoded as inline \texttt{<pause>} tokens within the transcript. Temporal fluency statistics (e.g., words/sec and pause count/duration) were also derived from the MFA alignments. Discourse labels were assigned via zero-shot inference with an LLM using the fixed cluster scheme, and ARPAbet phoneme sequences were extracted using HuPER~\cite{guo2026huper}. These views were aggregated into the JSON-structured prompt for each utterance.

\subsection{Experimental Setup}
 All models are fine-tuned from LLM checkpoints using the AdamW~\cite{loshchilov2017decoupled} optimizer with a cosine learning rate schedule, learning rate of $1\times10^{-4}$ and linear warm-up over the first 10\% of steps. LoRA~\cite{hu2022lora} is applied with rank $r=8$, $\alpha=16$ to the query and value projection matrices. We use participant turns as utterance units, following the timestamp-based segmentation provided by the dataset, and discard interviewer turns. Speaker-level predictions are aggregated via majority voting (dementia if more than half of the per-utterance predictions indicate dementia). Results are presented in terms of macro-averaged F1-score on the held-out test set.



\section{Results}

\subsection{Main Results}
Table~\ref{tab:main_results} compares our system with representative published approaches on ADReSSo~\cite{luz21_interspeech}. Prior systems emphasize different categories of speech-derived features, including temporal hesitation patterns, transcription fidelity, and multimodal fusion.  
  
The challenge baseline~\cite{luz21_interspeech} relies on conventional acoustic descriptors derived from the eGeMAPS~\cite{Eyben2016GeMAPS} feature set. WavBERT~\cite{zhu21e_interspeech} extends self-supervised acoustic embeddings with explicit pause-duration modeling, highlighting the diagnostic relevance of temporal hesitation cues. The Whisper-based system~\cite{Li24_whsiperbased} leverages a large-scale ASR backbone to obtain high-quality transcripts and speech representations, demonstrating that improved transcription fidelity can strengthen downstream diagnostic signals. Swin-BERT~\cite{Pan2024SwinBERT} further advances performance through a multimodal late-fusion architecture that combines separate audio and text encoders, representing the strongest prior system on this benchmark.  
  
Our framework integrates heterogeneous multi-view speech cues spanning duration-, phonological-, transcript-, and discourse-level information within a unified structured prompt. Despite its architectural simplicity, the model achieves the best performance on ADReSSo, surpassing the strongest prior system by a clear margin. Our approach consolidates speech-derived features into a unified reasoning process, enabling effective modeling through a single-pass inference without modality-specific encoders or late-stage feature aggregation.

To examine the role of model capacity, we further evaluate several smaller backbones. Performance improves consistently with model scale, suggesting that larger LLMs are better able to capture interactions among heterogeneous clinical cues. Nevertheless, even smaller models remain competitive with previously reported systems, indicating that the proposed multi-view reasoning framework is effective across a broad range of model sizes.

\begin{table}[t]
\centering
\small
\caption{Speaker-level F1-score (\%) on the ADReSSo test set.}\label{tab:main_results}
\begin{tabular}{@{}lc@{}}
\toprule
\textbf{System} & \textbf{F1-score } \\
\midrule
\multicolumn{2}{l}{\textit{Other published systems}} \\
Challenge baseline ~\cite{luz21_interspeech} & 78.92 \\
WavBERT~\cite{zhu21e_interspeech} & 83.10 \\  
Whisper-based~\cite{Li24_whsiperbased} & 84.50 \\ 
Swin-BERT~\cite{Pan2024SwinBERT} & 87.32 \\ 

\midrule
\multicolumn{2}{l}{\textit{Ours (multi-view LoRA)}} \\
Qwen3{-}4B  & 85.66 \\
Qwen3{-}8B  & 88.73 \\
Gemma3{-}12B{-}it & 88.72 \\
Qwen3{-}14B & \textbf{90.14} \\
\bottomrule
\end{tabular}
\end{table}

\subsection{Ablation Study}

Table~\ref{tab:ablation} reports a stepwise ablation study of our proposed model on ADReSSo. Starting from a speech-transcription-only baseline (81.48\% F1-score), each additional view yields a consistent improvement.
Topic and cluster labels (\textbf{+Topic \& cluster}, +5.81\%) provide the largest performance gain, highlighting the importance of discourse-level cues for dementia detection. By assigning each utterance to a predefined cluster and topic, the model obtains explicit signals about its semantic focus (e.g., scene content versus meta-discourse), allowing the model to better leverage discourse patterns characteristic of impaired descriptions.
Temporal fluency statistics (\textbf{+ Pause / duration}, +1.44\%) provide complementary global signals that augment the inline \texttt{<pause>} tokens. While pause tokens capture local disfluency patterns within the transcription, aggregated duration features summarize broader temporal characteristics such as speech rate and pause distribution, enabling the LLM to reason over higher-level fluency patterns beyond sequential context.
Phoneme sequences (\textbf{+Phoneme}, +1.41\%) capture sub-lexical articulatory patterns that are complementary to lexical content. These signals provide evidence of phonological irregularities that may not surface at the word level, confirming that phonological information contributes diagnostic value beyond transcript-based representations alone.

\begin{table}[t]
  \centering
  \caption{Ablation study on feature contributions (ADReSSo, speaker-level F1-score (\%)).}\label{tab:ablation}
  \begin{tabular}{@{}p{5.2cm}c@{}}
  \toprule
  \textbf{Feature set} & \textbf{F1-score} \\
  \midrule
  Speech transcription only                               & 81.48 \\
  \quad + Topic \& cluster                                & 87.29 \\
  \quad + Pause / duration\textsuperscript{\dag}          & 88.73 \\
  \quad + Phoneme                                         & \textbf{90.14} \\
  \bottomrule
  \multicolumn{2}{@{}p{6.2cm}@{}}{\footnotesize\textsuperscript{\dag} number of pauses, mean pause duration, words per second, \texttt{<pause>} tokens.}
  \end{tabular}
\end{table}

\section{Conclusion}

We presented a LoRA~\cite{hu2022lora}-tuned LLM framework for dementia detection that unifies four complementary speech-derived views—lexical transcripts, discourse-level cues, temporal fluency statistics, and phonological sequences—within a single structured prompt. Ablation experiments on ADReSSo confirm that each view contributes incremental diagnostic value, with discourse clusters providing the largest individual gain.

Performance scales consistently with model capacity (4B to 14B). Notably, even smaller models remain competitive with prior systems, suggesting that the multi-view formulation itself—rather than parameter scale alone—drives the observed improvements.

Overall, our results demonstrate that structured multi-view reasoning with parameter-efficient LLM adaptation offers a unified alternative to existing pipelines that rely on separate encoders or late-stage fusion. However, this study has several limitations. Extracting discourse-oriented representations relies on models accessed via commercial APIs, which restricts full control over the underlying training process. In addition, our evaluation is conducted only on English datasets, which may limit generalizability to other languages. Future work will extend this framework to multilingual benchmarks such as ADReSS-M~\cite{luz2024overview} and explore improving generalization by scaling-up both training corpora and model parameters. 

\section{Generative AI Use Disclosure}

We used generative AI to extract the discourse-oriented representation described in Section 2.2.3. GPT-5.2~\cite{openai2025gpt5} was consistently used for this extraction, and the instructions are provided in https://github.com/vivivic/is26dementia. Additionally, the core module for dementia detection, the LoRA-tuned LLM, is itself a generative AI model. We experimented with Qwen3~\cite{yang2025qwen3} and Gemma-3~\cite{gemmateam2025gemma3technicalreport} as base models.

\bibliographystyle{IEEEtran}
\bibliography{mybib}

\end{document}